\documentclass[a4paper,11pt]{article}
\usepackage{pos}
\usepackage{subcaption}
\usepackage{cleveref} 

\def\orcid#1{\kern .08em\href{https://orcid.org/#1}{\includegraphics[keepaspectratio,width=0.7em]{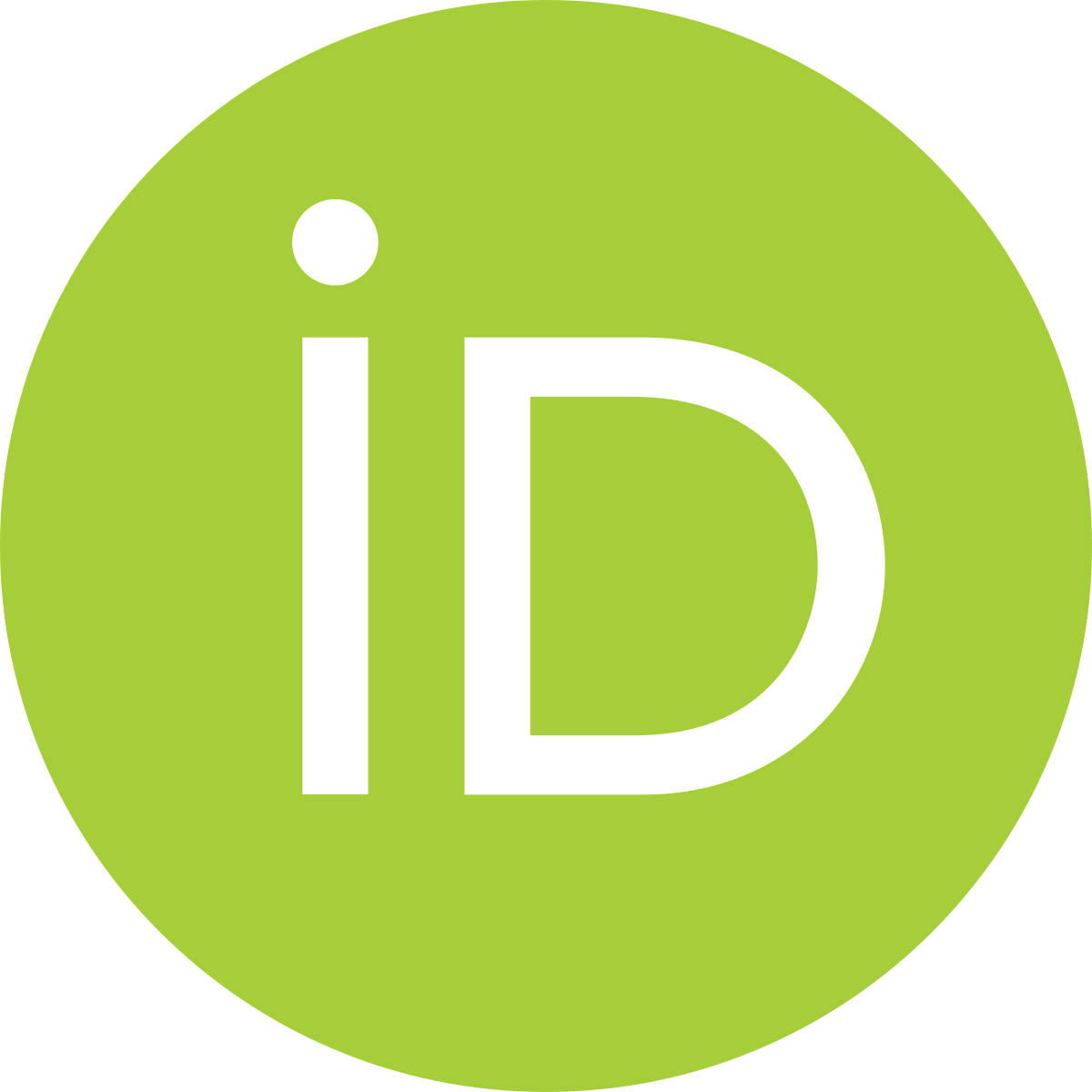}}}

\title{Analysis of hidden-charm pentaquarks as\\ triangle singularities via deep learning}
\ShortTitle{Analysis of hidden-charm pentaquarks as\\ triangle singularities via deep learning}

\author*{Darwin Alexander O. Co\orcid{0009-0003-8982-8727}}
\author{Denny Lane B. Sombillo\orcid{0000-0001-9357-7236}}

\affiliation{National Institute of Physics, University of the Philippines Diliman,\\
Quezon City 1101, Philippines}

\emailAdd{doco@up.edu.ph}
\emailAdd{dbsombillo@up.edu.ph}

\abstract{Identifying the nature of near-threshold enhancements is hindered by the limited resolution of experimental data leading to multiple conflicting interpretations. A prominent example of ambiguous line shape is the set of pentaquark signals observed by LHCb in 2019. Some of these signals can be interpreted as hadronic molecule, compact state, virtual state, or due to a kinematical triangle mechanism. In this work, we leverage the model-selection capability of deep neural networks to analyze and identify the nature of $P_{c\bar{c}}(4457)^+$. We trained a set of deep neural networks using line shapes with enhancements produced by triangle singularities and those produced by nearby poles. The training dataset for the triangle enhancements are generated by using a set of hadrons satisfying the required mass condition. The training line shapes for the pole-based classifications are generated using uniformized independent $S$-matrix poles configured to appear close to the relevant threshold. We found that, despite the presence of experimental uncertainties, the triangle mechanism is ruled out by the experimental data. The results also suggest that the data favor the pole-shadow pair interpretation for the $P_{c\bar{c}}(4457)^+$, which corresponds to a characteristic pole structure involving a resonance in a two-channel scattering system. Our result is consistent with the initial analysis done by LHCb favoring the Breit-Wigner fit over the triangle singularity. The present analysis offers an alternative approach to studying line shapes, supplementing the standard fitting methods.}

\FullConference{10th International Conference on Quarks and Nuclear Physics (QNP2024)\\
8-12 July, 2024\\
Barcelona, Spain\\}

\renewcommand{\printHeadAuthors}{DAO Co\,\orcid{0009-0003-8982-8727} and DLB Sombillo\,\orcid{0000-0001-9357-7236}}

\begin{document}
\maketitle

\section{Introduction}
Interpreting the nature of enhancements using line shapes from invariant mass distributions or scattering data is a challenging task in hadron spectroscopy. This difficulty arises primarily due to experimental uncertainties, which allow different, competing models to fit the same data equally well. Resolving ambiguity in line shape interpretation is crucial for accurately identifying which enhancements represent true unstable hadronic states. 

In \cite{Sombillo:2020ccg, Sombillo:2021ifs}, it was demonstrated that a neural network classifier can be trained to distinguish between threshold enhancements caused by a bound state pole and those produced by a virtual state pole. This result highlights the potential of using machine learning in analyzing enhancements observed in line shapes. Building on this, we explored the possibility of distinguishing pole-induced enhancements from kinematical enhancements in \cite{Co:2024bfl}. Specifically, we focused on the interpretation of the $P_{c\bar{c}}(4312)^+$ observed by LHCb in 2019 \cite{LHCb:2019kea}. Our analysis revealed that even with a relaxed physical parameter set in the training dataset, experimental data could still rule out the triangle singularity interpretation through machine learning-based line shape analysis. Furthermore, the study in \cite{Co:2024bfl} demonstrated that the $P_{c\bar{c}}(4312)^+$ strongly favors a pole-based interpretation, which was further investigated in \cite{Santos:2024}.

We now extend our framework to study the nature of the $P_{c\bar{c}}(4457)^+$, located near the $\Sigma_c^+\bar{D}^{*0}$ threshold. Unlike the $P_{c\bar{c}}(4312)^+$, the $P_{c\bar{c}}(4457)^+$ has a set of available hadrons that can participate in the triangle mechanism. In \cite{LHCb:2019kea}, it was argued that the signal is still best described by a Breit-Wigner (BW) parametrization; however, the triangle mechanism cannot yet be ruled out. Further scrutiny is required to clarify the nature of the $P_{c\bar{c}}(4457)^+$.

\section{TS vs. Pole-based line shapes}
In this study, we follow the framework discussed in \cite{Co:2024bfl}. For an analysis of the two-peak pentaquark structure $P_{c\bar{c}}(4440)^+$ and $P_{c\bar{c}}(4457)^+$, we restrict our region of interest to within 4400 to 4600 MeV and treat the nearby $\Sigma_c^+\bar{D}^{*0}$ as the second threshold. Note that the contributions of the lower $\Sigma_c^+\bar{D}^{0}$ threshold, close to the $P_{c\bar{c}}(4312)^+$, will be effectively parametrized as the background. Recall that the LHCb has also ruled out the triangle interpretation for the $P_{c\bar{c}}(4440)^+$ using similar reasoning for the $P_{c\bar{c}}(4312)^+$ \cite{LHCb:2019kea}. Hence, we parametrize this structure using a Breit-Wigner, as suggested by similar studies, e.g. \cite{Nakamura:2021qvy}, and explore the triangle singularity of the $P_{c\bar{c}}(4457)^+$.

We generated the training dataset for the triangle singularity line shape using the amplitude given in \eqref{eq:TS-Int}. The singularities of the integrand determine the conditions under which an enhancement appears in the invariant mass distribution $m_{23}$ of the final state. Specifically, if the mass of the first intermediate particle $m_1$ lies within a specific range, a peak is observed in $m_{23}$ within the corresponding interval, as described in \eqref{eq:mass_range}. In principle, the triangle mechanism is always accompanied by hadron interaction that leads to the final state, which means that enhancement due to pole can still be present. To remove this possibility, the effective meson-baryon interaction connecting the intermediate 23 particles to the final state C is modeled as a separable-type interaction with a Yamaguchi form factor \cite{Yamaguchi:1954mp}. We made the interaction relatively weak to ensure that the enhancement is dominated by the triangle singularity and not by poles, which is done by setting the cut-off parameter to be within the range of 2000 to 3000 MeV. 

\begin{equation}
    I(k)\propto \int_0^\infty \frac{q^2\,dq}{P^0-\sqrt{m_1^2+q^2}-\sqrt{m_2^2+q^2}+i\epsilon}
    \int_{-1}^{1} \frac{dz}{E_C-\omega(q)-\sqrt{m_3^2+q^2+k^2-2qkz}+i\epsilon}
    \label{eq:TS-Int}
\end{equation}
\begin{equation}
    \resizebox{0.92\textwidth}{!}{$m_1^2 \in \left[\frac{M_A^2m_3+M_B^2m_2}{m_2+m_3}-m_2m_3,\;(M_A-m_2)^2\right],
    \;
    m_{23}^2 \in \left[(m_2+m_3)^2,\;\frac{M_Am_3^2-M_B^2m_2}{M_A-m_2}+M_Am_2\right]$}\\[5pt]
    \label{eq:mass_range}
\end{equation}

We consider three relevant triangle loops for $P_{c\bar{c}}(4457)^+$ which we call Triangle A, B, and C: $D_{s1}^-\Lambda_c(2595)^+\bar{D}^0$, $\Lambda^*\chi_{c0}p$, and $\Lambda^*\chi_{c1}p$, where we set $m_1$ to be within the range of $[2832,2886]$, $[2162,2205]$, and $[1900,2050]$ MeV, respectively, such that an enhancement will be produced within our region of interest, i.e. above the $P_{c\bar{c}}(4440)^+$ BW peak, but not beyond 4600 MeV. The rest of the masses follow the data reported by PDG, while the width $\epsilon$ is randomly selected from 1.0 to 10.0 MeV in order to simulate prominent peaks. Note that the possible interference between the BW and TS was also accounted for in the dataset.

The training amplitudes with pole enhancements are generated using the uniformized $S$-matrix \cite{Santos:2023gfh,Kato:1965iee}. A uniformized variable $\omega$ is introduced such that $\omega=(q_1+q_2)/\sqrt{\epsilon_2^2-\epsilon_1^2}$, where $q_k$ is the two-hadron center of mass momentum and $\epsilon_k$ is the threshold of the $k^{\text{th}}$ channel. To fully utilize the two-channel parametrization, we choose the $J/\psi p$ as channel 1 and $\Sigma_c\bar{D}^{*0}$ as channel 2. The $S$-matrix is parametrized using the independent set of poles $\{\omega_{m}\}_{m=1}^M$ such that the relevant element of our two-channel S-matrix is 
\begin{equation}
    S_{11}(\omega)=\prod_{m=1}^M
    \dfrac{D_m(1/\omega)}{D_m(\omega)};\quad
    D_m(\omega)=\dfrac{1}{\omega^2}
    (\omega-\omega_m)(\omega+\omega_m^*)
    (\omega-\omega_{\bar{m}})(\omega+\omega_{\bar{m}}^*)
    \label{eq:jost}
\end{equation}
with $|\omega_m\omega_{\bar{m}}|=1$. Poles were produced on different Riemann sheets within the range $Re(E_{m})\in[\epsilon_2-10,4500]$ MeV with $Im(E_{m})\in[0,50]$ MeV, following the pole-counting argument \cite{Morgan:1992}.

\begin{figure}[ht!]
    \centering
    \includegraphics[width=0.85\linewidth]{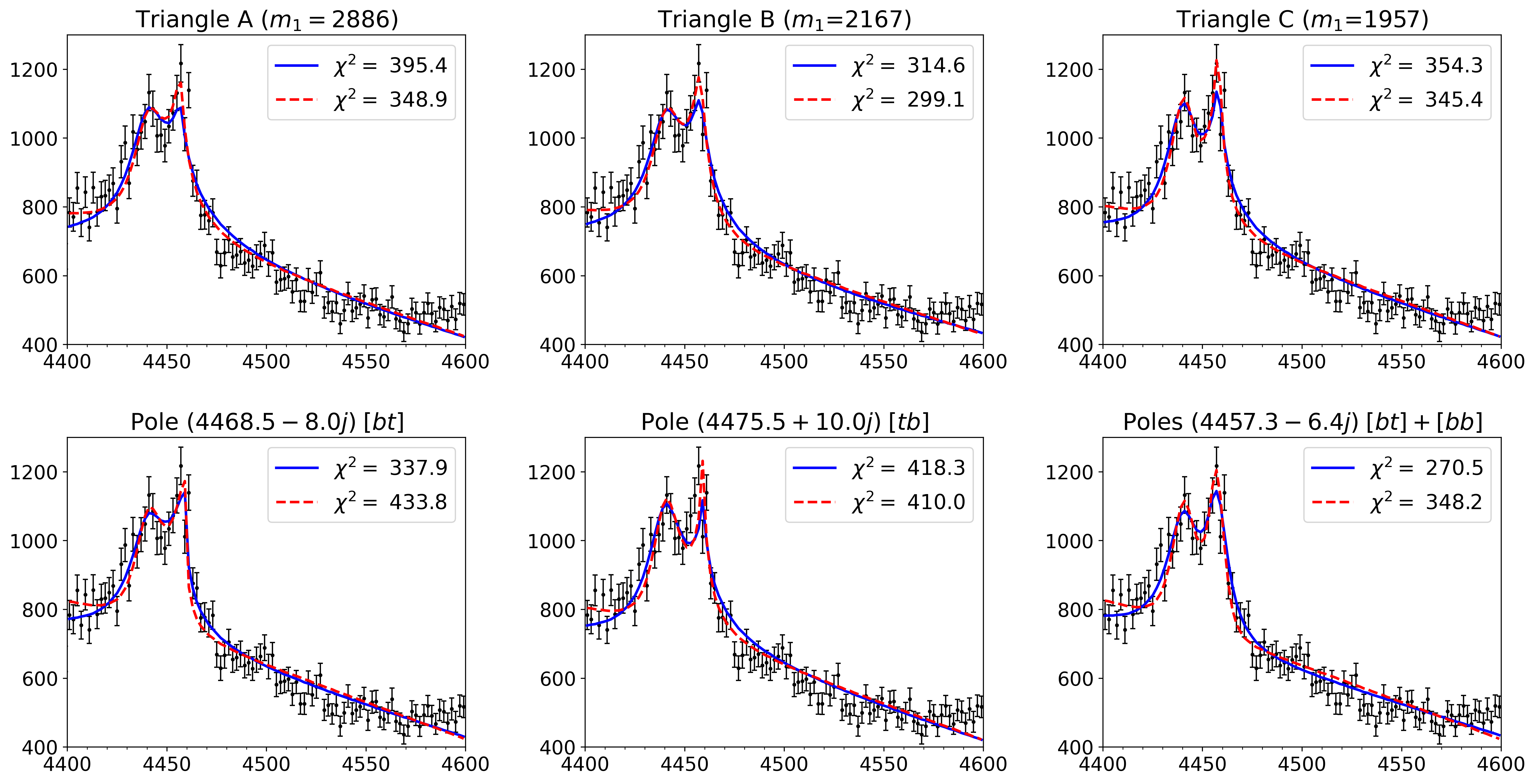}
    \caption[]{Best-fit amplitudes for the two-peak pentaquark structure plotted over LHCb experimental data (retrieved from \cite{hepdata.89271}). A Breit-Wigner was used for the $P_{c\bar{c}}(4440)^+$ structure, while different mechanisms were considered for $P_{c\bar{c}}(4457)^+$ as indicated in the titles. The red lines account for interference of the two.}
    \label{fig:curvefit}
\end{figure}

To illustrate the line-shape ambiguity, a simple curve fitting and standard Chi-square test were performed using the two formalisms discussed above. We observe in \Cref{fig:curvefit} that Triangle B and the pole-shadow pair configuration ($[bt] + [bb]$) achieved the best yet comparable $\chi^2$-scores. Since these amplitudes involved the same number of free parameters, it is difficult to argue which model is better in describing the data. This motivates us to employ the deep learning framework.

\section{DNN design and training}
\begin{figure}[ht!]
    \centering
    \begin{subfigure}[t]{0.49\textwidth}
        \centering
        \includegraphics[width=1\linewidth]{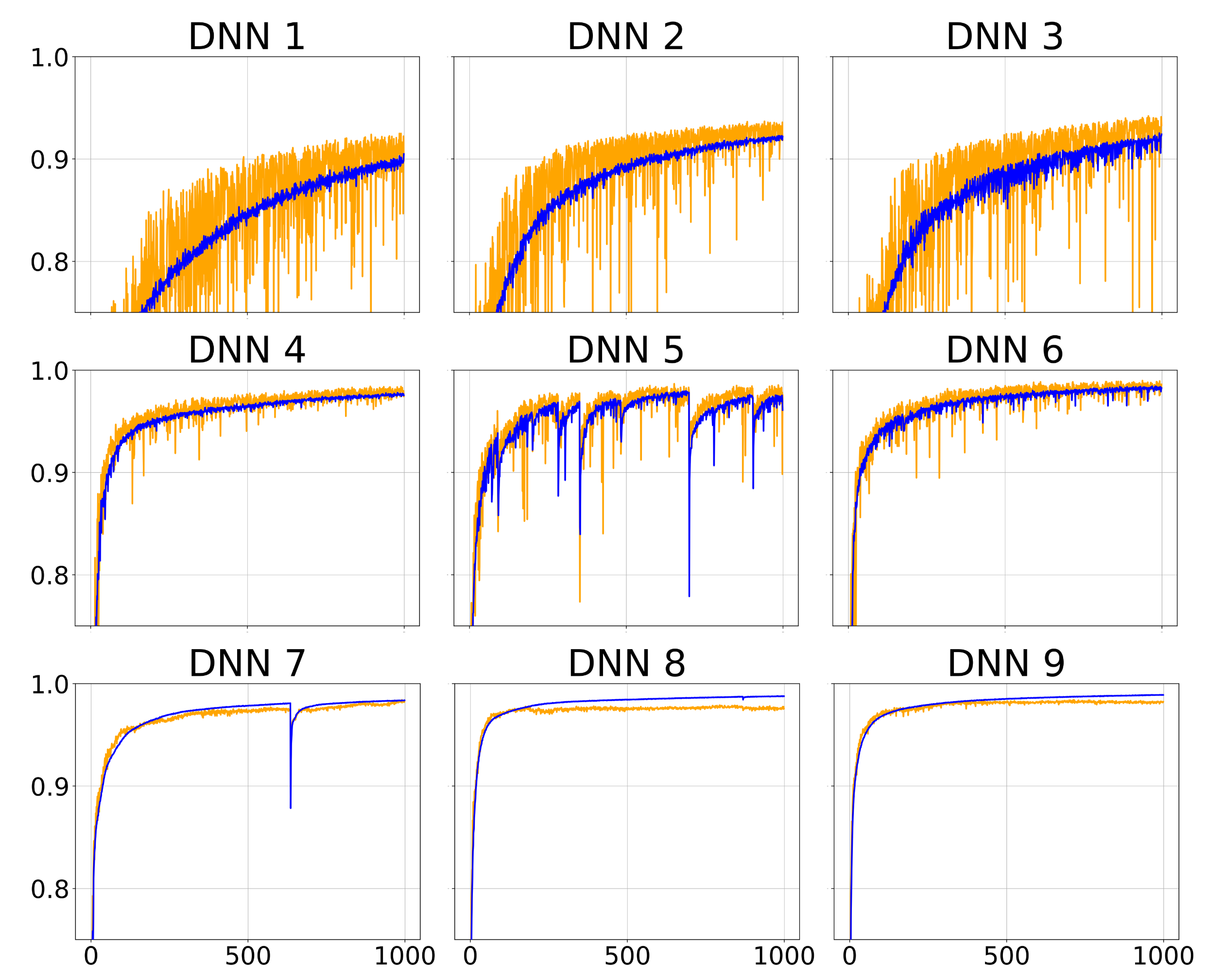}
        \caption[]{ }\label{fig:accu}
    \end{subfigure}\,
    \begin{subfigure}[t]{0.49\textwidth}
        \centering
        \includegraphics[width=1\linewidth]{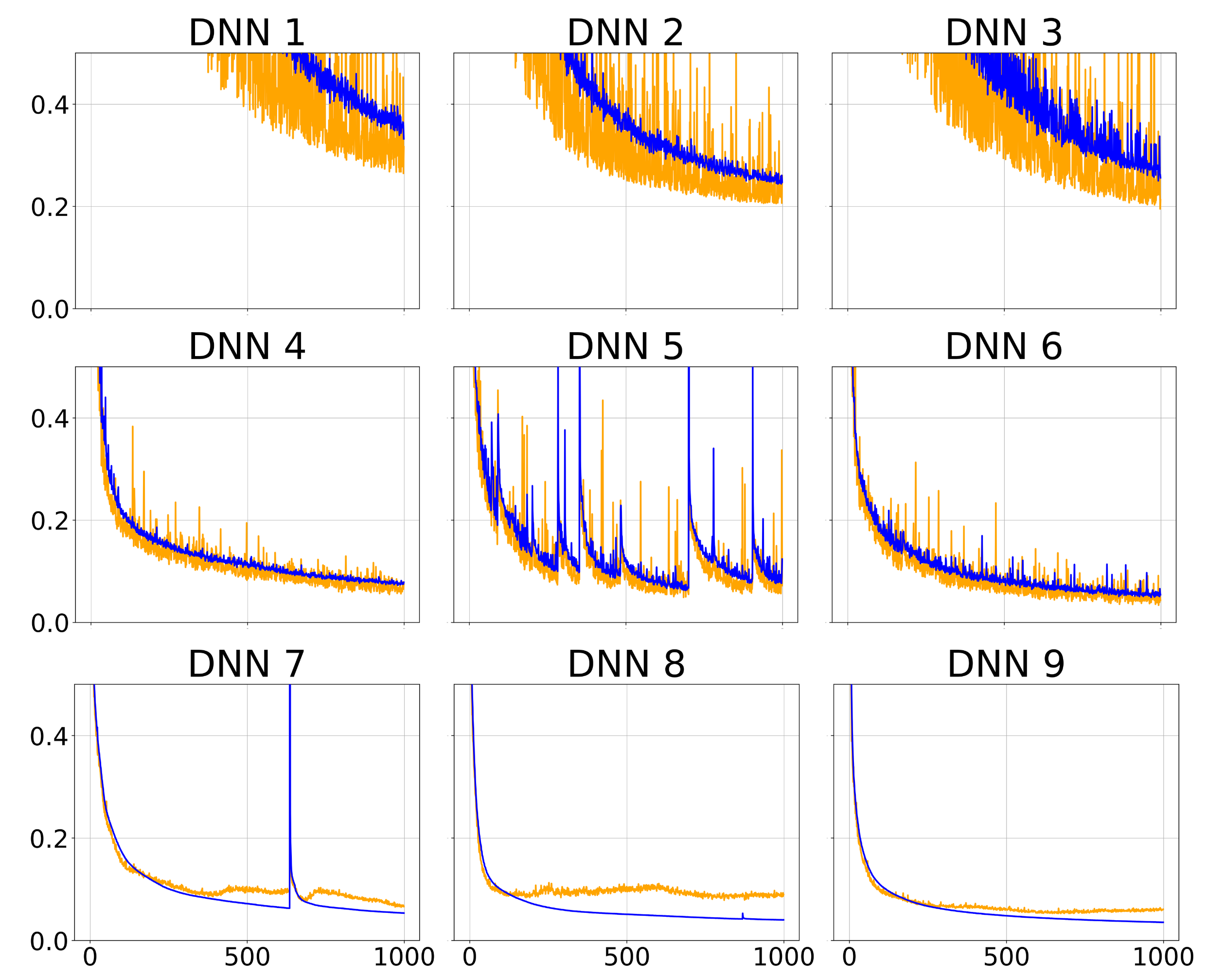}
        \caption[]{ }\label{fig:loss}
    \end{subfigure}
    \caption[]{Training performance of DNN models (as listed in \Cref{tab:dnnmodels}): (a) accuracy per epoch, (b) loss per epoch. Blue lines are for the training dataset, while orange lines are for the testing set.}
    \label{fig:DNN-perf}
\end{figure}

We construct our deep neural network (DNN) with 200 input nodes that accept energy and intensity values forming each line shape and 4 output nodes each corresponding to an interpretation (i.e. triangle singularity, 1 pole in $[bt]$, 1 pole in $[tb]$, and 1 pole each in $[bt]$ and $[bb]$).
A total of $120,000$ datasets were generated and used in the training. Multiple DNN models were constructed to find the optimal design by considering different hidden-layer architectures and optimizers, as listed in \Cref{tab:dnnmodels}. We find that all 9 models achieved good overall performance after 1000 training epochs, with final accuracy greater than $0.90$ and loss lower than $0.40$, as shown in \Cref{fig:DNN-perf}.

\begin{table}[ht!]
    \centering
    {\renewcommand{\arraystretch}{1.05}
    \begin{tabular}{c@{\hskip 0.2in}l}
    \hline\hline
    \textbf{DNN model}& \textbf{Optimizer and architecture} \\
    \hline
        DNN 1& AdaGrad: 200-[350-250]-4  \\
        DNN 2& AdaGrad: 200-[350-300-250]-4 \\
        DNN 3& AdaGrad: 200-[350-400-350]-4 \\
        \hline
        DNN 4& AMSGrad: 200-[350-250]-4  \\
        DNN 5& AMSGrad: 200-[350-300-250]-4 \\
        DNN 6& AMSGrad: 200-[350-400-350]-4 \\
        \hline
        DNN 7& SMORMS3: 200-[350-250]-4  \\
        DNN 8& SMORMS3: 200-[350-300-250]-4 \\
        DNN 9& SMORMS3: 200-[350-400-350]-4 \\
    \hline\hline
    \end{tabular}}
    \caption{List of trained DNN models with corresponding optimizer and architecture.}
    \label{tab:dnnmodels}
\end{table}

\section{Model selection results}
\begin{figure}[ht!]
    \centering
    \includegraphics[width=0.55\linewidth]{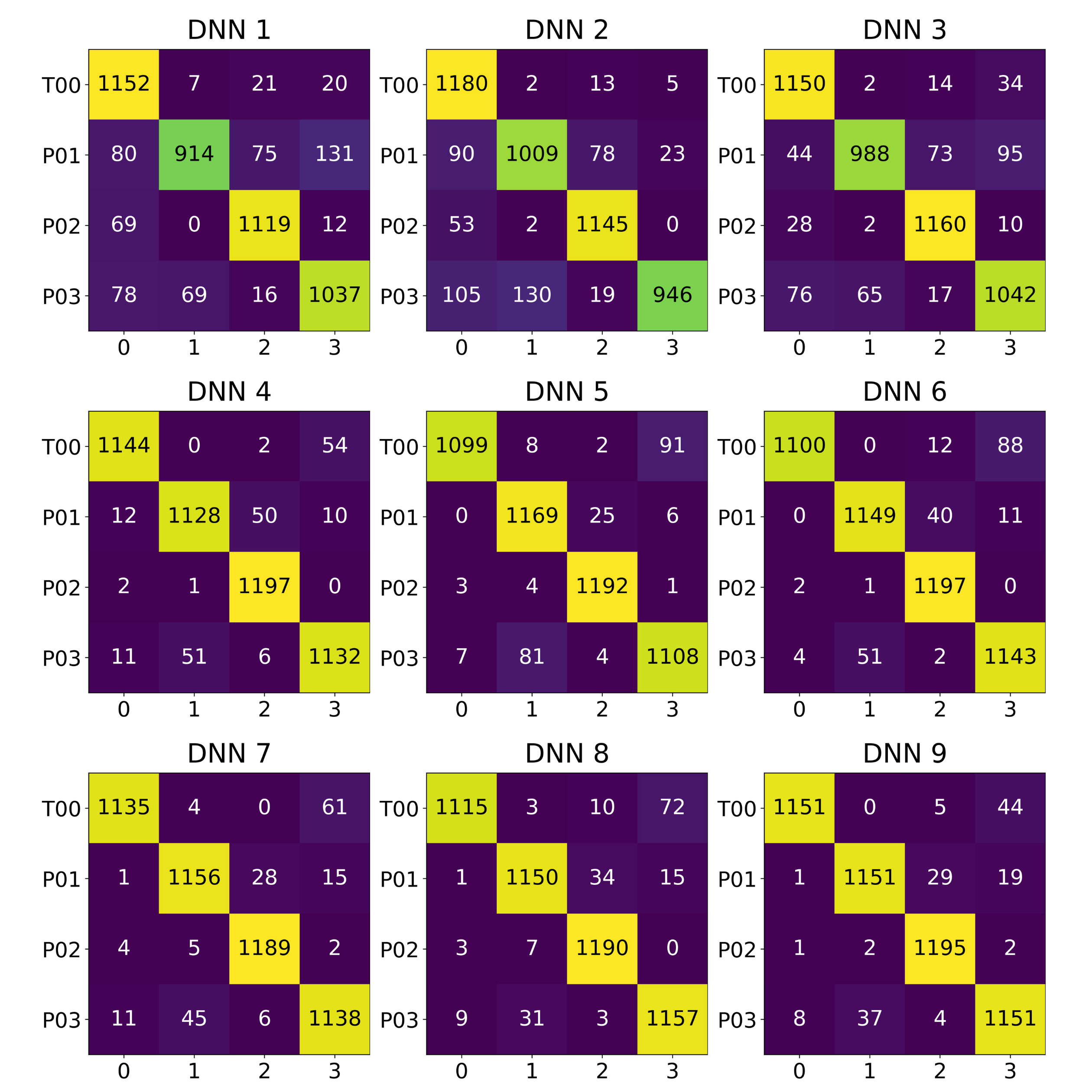}
    \caption[]{Confusion matrix of each trained DNN model. T00, P01, P02, and P03 are the true labels while 0, 1, 2, and 3 are the predicted labels.}
    \label{fig:confusion}
\end{figure}

We further analyzed the performance of our trained DNN models by investigating their respective confusion matrices. These were constructed using a separately generated set of 1200 validation data for each class. We see in \Cref{fig:confusion} that the DNN's can make predictions correctly as evidenced by a large percentage in the diagonal elements, but minimal misclassifications are still present in the off-diagonal elements, especially in DNN's 1-3.

We now use our trained and validated DNN models to analyze the LHCb data within the chosen energy range. A total of 3000 line shapes were constructed using bootstrap technique by exploiting the statistical uncertainties in the intensity, while also randomly choosing as representative value for each energy bin. In order to produce some variations in the inference, we adopted the snapshot ensemble method. We retrieve the previous training states of our DNN and feed the bootstrapped line shapes. The final inference results with reported mean and standard dev. is shown in \Cref{tab:inference}. The results show that the experimental data largely favors the pole-shadow pair interpretation over the triangle interpretation for the $P_{c\bar{c}}(4457)^+$, which is a characteristic of a true compact resonance.

\begin{table*}[ht!]
    \centering
    {\renewcommand{\arraystretch}{1.05} \begin{tabular*}{\textwidth}{l@{\hskip 0.2in}c@{\hskip 0.2in}c@{\hskip 0.2in}c@{\hskip 0.2in}c}
    \hline\hline
    &\textbf{Triangle} & \textbf{1 pole in $[bt]$} & \textbf{1 pole in $[tb]$} & \textbf{1 pole each in $[bt]$ \& $[bb]$} \\
    \hline\hline
    DNN 1 
    &$1.55 \pm 2.36$ &$2387.00 \pm 246.31$ 
    &$0.90 \pm 0.30$ &$610.55 \pm 244.91$ \\
    DNN 2
    &$0 \pm 0$ &$1.63 \pm 1.64$ 
    &$0 \pm 0$ &$2998.37 \pm 1.64$ \\
    DNN 3
    &$0 \pm 0$ &$1.92 \pm 3.40$ 
    &$0.63 \pm 0.70$ &$2997.45 \pm 3.62$ \\
    \hline\hline
    DNN 4
    &$188.63 \pm 135.30$ &$1141.25 \pm 292.44$ 
    &$0 \pm 0$ &$1670.12 \pm 332.91$ \\
    DNN 5
    &$1.63 \pm 9.22$ &$11.80 \pm 73.53$ 
    &$0 \pm 0$ &$2986.57 \pm 73.85$ \\
    DNN 6
    &$0 \pm 0$ &$8.05 \pm 8.66$ 
    &$0 \pm 0$ &$2991.95 \pm 8.66$ \\
    \hline\hline
    DNN 7
    &$76.50 \pm 92.76$ &$164.30 \pm 48.84$ 
    &$0 \pm 0$ &$2759.20 \pm 55.63$ \\
    DNN 8
    &$127.67 \pm 14.02$ &$50.45 \pm 5.80$ 
    &$30.60 \pm 5.57$ &$2791.28 \pm 14.13$ \\
    DNN 9
    &$0.10 \pm 0.30$ &$26.25 \pm 2.96$ 
    &$0 \pm 0$ &$2973.65 \pm 2.93$ \\
    \hline\hline
  \end{tabular*}}
  \caption{Final inference results using LHCb experimental data. The table shows the DNN predictions: mean count out of 3000 bootstrapped inference amplitudes and their corresponding standard deviations obtained via the snapshot ensemble approach for 40 instances of each DNN model.}
  \label{tab:inference}
\end{table*}

\section{Conclusion and Outlook}
In this work, we scrutinized the possible triangle interpretation of $P_{c\bar{c}}(4457)^+$ as suggested in \cite{LHCb:2019kea}. We performed a pure line shape analysis by training a set of DNN's that act as model-selection maps. Our result is consistent with the conclusion that a Breit-Wigner line shape is a more plausible explanation of the $P_{c\bar{c}}(4457)^+$, thereby ruling out the triangle singularity interpretation despite the presence of experimental uncertainties. It remains to be determined whether the signals are produced by the double triangle mechanism as pointed out in \cite{Nakamura:2021qvy}. This alternative interpretation should be included in future analysis. 

\bibliographystyle{JHEP}
\bibliography{mybib}

\providecommand{\href}[2]{#2}\begingroup\raggedright\begin{thebibliography}{10}

\bibitem{Sombillo:2020ccg}
D.L.B.~Sombillo, Y.~Ikeda, T.~Sato and A.~Hosaka, \href{https://doi.org/10.1103/PhysRevD.102.016024}{\emph{Phys. Rev. D} {\bfseries 102} (2020) 016024} [\href{https://arxiv.org/abs/2003.10770}{{\ttfamily 2003.10770}}].

\bibitem{Sombillo:2021ifs}
D.L.B.~Sombillo, Y.~Ikeda, T.~Sato and A.~Hosaka, \href{https://doi.org/10.1007/s00601-021-01642-z}{\emph{Few Body Syst.} {\bfseries 62} (2021) 52} [\href{https://arxiv.org/abs/2106.03453}{{\ttfamily 2106.03453}}].

\bibitem{Co:2024bfl}
D.A.O.~Co, V.A.A.~Chavez and D.L.B.~Sombillo,  [\href{https://arxiv.org/abs/2403.18265}{{\ttfamily 2403.18265}}].

\bibitem{LHCb:2019kea}
{\scshape {LHCb}} collaboration, \href{https://doi.org/10.1103/PhysRevLett.122.222001}{\emph{Phys. Rev. Lett.} {\bfseries 122} (2019) 222001} [\href{https://arxiv.org/abs/1904.03947}{{\ttfamily 1904.03947}}].

\bibitem{Santos:2024}
L.M.~Santos, V.A.~Chavez and D.L.B.~Sombillo, \href{https://doi.org/10.1088/1361-6471/ad8ee3}{\emph{J. Phys. G: Nucl. Part. Phys.} (2024) } [\href{https://arxiv.org/abs/2405.11906}{{\ttfamily 2405.11906}}].

\bibitem{Nakamura:2021qvy}
S.X.~Nakamura, \href{https://doi.org/10.1103/PhysRevD.103.L111503}{\emph{Phys. Rev. D} {\bfseries 103} (2021) 111503} [\href{https://arxiv.org/abs/2103.06817}{{\ttfamily 2103.06817}}].

\bibitem{Yamaguchi:1954mp}
Y.~Yamaguchi, \href{https://doi.org/10.1103/PhysRev.95.1628}{\emph{Phys. Rev.} {\bfseries 95} (1954) 1628}.

\bibitem{Santos:2023gfh}
L.M.~Santos and D.L.B.~Sombillo, \href{https://doi.org/10.1103/PhysRevC.108.045204}{\emph{Phys. Rev. C} {\bfseries 108} (2023) 045204} [\href{https://arxiv.org/abs/2308.03325}{{\ttfamily 2308.03325}}].

\bibitem{Kato:1965iee}
M.~Kato, \href{https://doi.org/10.1016/0003-4916(65)90235-6}{\emph{Annals Phys.} {\bfseries 31} (1965) 130}.

\bibitem{Morgan:1992}
D.~Morgan, \href{https://doi.org/10.1016/0375-9474(92)90550-4}{\emph{Nucl. Phys. A} {\bfseries 543} (1992) 632}.

\bibitem{hepdata.89271}
{\scshape {LHCb}} collaboration, {HEPData (collection)}, 2019.
\newblock \url{https://doi.org/10.17182/hepdata.89271}.

\end{thebibliography}\endgroup

\end{document}